\documentclass[twocolumn,floatfix,pra,showpacs]{revtex4}
\usepackage{graphicx}
\usepackage{amssymb}
\usepackage{amsmath}
\usepackage{color}
\usepackage{psfrag}
\usepackage{epsfig}
\usepackage{bbm}
\usepackage{bm}
\newcommand{\ket}[1]{| #1 \rangle}

\newcommand{\rb}[1]{\left( #1 \right)}

\newcommand{\ew}[1]{\langle #1 \rangle}
\newcommand{\beq}{\begin{eqnarray}}
\newcommand{\eeq}{\end{eqnarray}}

\newcommand{\op}[2]{| #1 \rangle \langle #2 |}

\newcommand{\eq}[1]{Eq.~(\ref{#1})}
\newcommand{\fig}[1]{Fig.~\ref{#1}}

\newcommand{\kett}[1]{| #1 \rangle\!\rangle }
\newcommand{\braa}[1]{\langle\!\langle #1|}
\newcommand{\eww}[1]{\langle\! \langle #1\rangle\! \rangle}
\newcommand{\opp}[2]{| #1 \rangle\! \rangle\langle\! \langle #2 |}
\begin{document}
\title{Counting statistics of cotunneling electrons}
\author{Clive Emary}
\affiliation{
  Institut f\"ur Theoretische Physik,
  Technische Universit\"at Berlin,
  D-10623 Berlin,
  Germany
}

\date{\today}
\begin{abstract}
We describe a method for calculating the counting statistics of electronic transport through nanoscale devices with both  sequential and cotunneling contributions.  The method is based upon a perturbative expansion of the von Neumann equation in Liouvillian space, with current cumulants calculated from the resulting  nonMarkovian master equation without further approximation.  As application, we consider transport through a single quantum dot and discuss the effects of cotunneling on noise and skewness, as well as the properties of various approximation schemes.
\end{abstract}
\maketitle


Cotunneling, the transfer of electrons via intermediate ``virtual'' states, can be an important mechanism in the transport of electrons through quantum dots (QDs) \cite{ave90,ave92}.  In the Coulomb blockade regime,  sequential tunnelling processes are exponentially suppressed and, since it only suffers an algebraic suppression, cotunneling becomes the dominant current-carrying mechanism.  Experimental interest in cotunneling has remained high from the earliest experiments on metallic grains \cite{gee90} and large quantum dots \cite{pas93}, through to more modern experiments on few-electron single- \cite{fra01, zum04,sch05} and double- \cite{sig06,gus08} quantum dots. 

From a theoretical perspective, cotunneling refers to processes fourth-order in the coupling between the system and the leads.   Such processes can be taken into account in a number of different ways, e.g. \cite{ave92,suk01,gol04, ped05,sch94,kon96,thi03,thi05}.  Most relevant here is the real-time diagrammatic approach\cite{sch94,kon96,thi03,thi05}, in which higher-order tunnelling processes are incorporated into a master equation in a systematic fashion .  This theory has been extensively developed and successfully applied to numerous transport problems: not just single QDs, but also double dots \cite{wey08DQD}, quantum dot spin-valves \cite{bra04}, carbon nanotubes \cite{wey08CNT}, and QD-interferometers \cite{urb08}.
Whilst such higher-order calculations have typically been restricted to the stationary current, and more recently, the shotnoise \cite{thi03,thi05}, much interest presently surrounds the full counting statistics (FCS) of the current, i.e. in current correlations beyond the second-order shotnoise \cite{lev93,yul99,bag03}. The last few years has seen the advent of experiments capable of detecting the passage of single electrons through QD systems \cite{fuj06,gus06,Sukh07,fri07,fli09} and the experimental determination of FCS. Recently, measurements of the 15th cumulant were reported for a single quantum dot\cite{fli09}.

In this article we bring together several strands to investigate the influence of cotunneling on FCS.  We derive a fourth-order master equation for the reduced density matrix of an arbitrary mesoscopic system using the Liouvillian-space perturbation theory of Ref.~\cite{lei08} and show how counting fields may be added in this formalism.
Given that the resulting master equation is nonMarkovian, we employ the nonMarkovian formalism of Flindt {\it et al.} \cite{fli08} to obtain expressions for the current cumulants. 

The calculation of the FCS for a nonMarkovian ME with cotunneling was described in Ref. \cite{bra06}, but the approach followed here is somewhat different.  Braggio {\it et al.} \cite{bra06} calculate the cumulant generating function (CGF), and thus all the cumulants, rigorously to fourth order in the dot-lead coupling.  Here, we calculate the Liouvillian rigorously to fourth-order and make no further approximations in obtaining the cumulants. It is one of the aims of this paper to investigate the differences between the predictions of these two approaches.

We use the transport through a single QD as our example system.  We study first the single resonant level (SRL) model.  Exact solutions exist for this model and this allows an evaluation of various approximation schemes.  We also study the effects of interaction on transport with an Anderson model.  Of the higher-order cumulants, we focus here on the skewness as the first correlator beyond the shotnoise. We compare, both on a formal and a numerical level, with the work of Braggio {\it et al} \cite{bra06}, and with the shotnoise results of Thielmann {\it et al} \cite{thi05}.

\section{Transport Model}

We begin by specifying the general transport set-up under consideration here.  The total Hamiltonian 
$
  H = H_\mathrm{res} + H_\mathrm{S} + V
$
is composed of reservoir, system, and interaction parts.  We write the system part in its diagonal basis 
$ H_\mathrm{S} = \sum_a E_a \op{a}{a}$, where $\ket{a}$ is a many-body system state of $N_a$ electrons.
We consider a set of reservoirs, labelled with an index $\alpha$ that includes spin and any other relevant quantum numbers.  We assume noninteracting reservoirs with Hamiltonian
\beq
  H_\mathrm{res} = \sum_{k,\alpha} (\omega_{k\alpha} +\mu_\alpha)
  a^\dag_{k\alpha} a_{k\alpha}
  ,
  \label{Hres}
\eeq
where $\omega_{k\alpha}$ is the energy of the $k$th mode in lead $\alpha$, $a_{k\alpha}$ is a lead annihilation operator, and we have included the chemical potential of lead $\alpha$, $\mu_\alpha$, at this point for convenience.

To ease book-keeping, we introduce a compact single index ``$1$'' to denote the triple of indices $(\xi_1,k_1,\alpha_1)$.  The first index $\xi_1=\pm$ indicates whether a reservoir operator is a creation or annihilation operator:
\beq
  a_1 =  a_{\xi_1 k_1 \alpha_1}
  =
  \left\{ 
    \begin{array}{c c}
      a^\dag_{k_1\alpha_1}, &\quad \xi_1 =+\\
      a_{k_1\alpha_1}, &\quad \xi_1 =-
    \end{array}
  \right. 
  .
\eeq
Leaving sums implicit, the reservoir Hamiltonian reads
\beq
  H_\mathrm{res} &=&\rb{ \omega_{k \alpha}+\mu_\alpha} a_{+k\alpha} a_{-k\alpha}
  \nonumber \\
  &=& 
  \rb{ \omega_1+\mu_1} a_1 a_{\overline{1}}\delta_{\xi_1,+}
  .
\eeq
In equilibrium, the reservoir electrons are distributed according to the Fermi function
\beq
   f(\omega) = \frac{1}{e^{\omega/k_B T} +1}
   ,
\eeq
which, since we include the chemical potential in \eq{Hres} and assume a uniform temperature, is the same for all reservoirs.

Single-electron tunnelling between system and reservoirs is described by the Hamiltonian
\beq
  V =  \sum_{k \alpha m} 
  t_{k \alpha m} a^\dag_{k\alpha} d_m 
  + t^*_{k \alpha m} d^\dag_m a_{k\alpha}
  \label{V1}
  ,
\eeq
where $d_m$ is the annihilation operator for single-particle level $m$ in the system, and $t_{k \alpha m}$ is a tunnelling amplitude. 
We write this interaction as
\beq
  V&=&  \xi_1 t_{1m} a_1 j_{\xi_1 m}
  \label{Vtj}
  ,
\eeq
with coefficients 
$t_{+ k \alpha m} = t_{k \alpha m}$ and
$t_{- k \alpha m} = t^*_{k \alpha m}$,
and system operators in the many-body system basis
\beq
  j_{+ m} = \sum_{aa'} \ew{a |d_m | a'} \delta\rb{N_a-N_{a'}+1} \op{a}{a'}
  ,
\eeq
and $j_{-  m} = j_{ m}^\dag$. We have made explicit here the change in system charge induced by the operator. Although these operators only depend on $\xi_1$, we label them with the full ``$1$'' index for convenience: $j_{1 m} = j_{\xi_1 m}$.

At time $t=0$ we posit a separable total density matrix:
\beq
  \rho(t=0) = \rho_\mathrm{S}(0) \rho_\mathrm{res}^\mathrm{eq}
  ,
\eeq
with the system in arbitrary state $\rho_\mathrm{S}(t_0)$ and reservoirs in thermal equilibrium.

\section{Liouville-Laplace space}

We now construct the elements required to perform our perturbation calculation in Liouville-Laplace space. In this section and the next, we follow Refs.\cite{lei08,kor07,sch08}.
The total density matrix evolves according to the von Neumann equation:
\beq
  \dot{\rho}(t) = -i \left[H,\rho(t)\right] =  {\cal L} \rho(t),
  \label{rhodot}
\eeq
which defines the Liouvillian super-operator $ {\cal L} = -i\left[H,\bullet~\right]$.  
This Liouvillian consists of three parts:
\beq
  {\cal L} = {\cal L}_\mathrm{res} + {\cal L}_\mathrm{S} + {\cal L}_V
\eeq
with ${\cal L}_\mathrm{res}= -i\left[H_\mathrm{res},\bullet~\right]$,  ${\cal L}_\mathrm{S}=-i\left[H_\mathrm{S},\bullet~\right]$, and 
$ {\cal L}_V = -i\left[V,\bullet~\right]$.
We write the interaction Liouvillian as
\beq
{\cal L}_V =- i \xi_1 t_{1m} \sum_p  A_1^p J_{1 m}^p
  ,
\eeq
where $p=\pm$ is a Keldysh index corresponding to the two parts of the commutator. Superoperators $A$ and $J$ are defined through their actions on arbitrary operator $O$: For the reservoir, we have
\beq
  A^p_1 O = 
  \left\{
   \begin{array}{c c}
      a_1 O, &p =+ \\
     O a_1, &p =-
    \end{array}
  \right.
  \label{ddef}
  ,
\eeq
and analogously for the system
\beq
  J^p_{1m} O = 
  \left\{
   \begin{array}{c c}
      j_{1m} O, &p =+ \\
      O j_{1m}, &p =-
    \end{array}
  \right.
  \label{Jdef}
  .
\eeq

By organising the elements of density matrices into vectors, superoperators, such as the Liouvillian, take the form of matrices.  This is a particularly convenient representation for the system Liouvillian. We write a general system density matrix, $\rho_S = \sum_{a_1a_2} \rho_{a_1 a_2} \op{a_1}{a_2}$,  as the  vector $\kett{\rho_S} = \sum_{a} \rho_{a} \kett{\phi_a}$, where the single index $a$ corresponds to the double $(a_1,a_2)$ such that the ``ket'' $\kett{\phi_a}$ corresponds to $\op{a_1}{a_2}$.
The action of the free system Liouvillian $L_S$ on vector $\kett{\phi_a}$ is
\beq
  L_S \kett{\phi_a} \equiv -i \left[H_S, \op{a_1}{a_2} ~\right] 
  = -i \Delta_{a}\kett{\phi_a}
  ,
\eeq
where $\Delta_{a} \equiv E_{a_1} - E_{a_2}$ defines the Bohr frequencies.  The vectors $\kett{\phi_a}$ are therefore the right eigenvectors of Liouvillian $L_S$. The  left eigenvectors, $\braa{\phi_a}$, fulfil
\beq
\braa{a} L_S = -i \Delta_{a}\braa{\phi_a} 
\eeq
and together with the right eigenvectors form a bi-orthonormal set:
$
  \eww{\phi_a|\phi_{a'}} = \delta_{a,a'}
$. 
We have the completeness relation in Liouvillian space
\beq
  \mathbbm{1} = \sum_{a}\opp{\phi_a}{\phi_a} 
  \label{cset}
  .
\eeq
In general, it is important to make the distinction between left and right eigenvectors because an arbitrary super-operator, in particular the effective system Liouvillian, will not be Hermitian, and the left and right eigenvectors are therefore not adjoint.

\section{Effective Liouvillian}
With the definition of the Laplace transform
\beq
  \rho(z) \equiv \int_{0}^\infty dt e^{-z t} \rho(t)
  ,
\eeq
equation (\ref{rhodot}) yields the solution
\beq
  \rho(z) = \frac{1}{z-{\cal L}} \rho(0)
  .
  \label{rhozfull}
\eeq
Tracing \eq{rhozfull} over reservoir degrees of freedom, results in an expression for the reduced density matrix of the system that we write
\beq
  \rho_\mathrm{S}(z) = \frac{1}{z- {\cal W}(z) } \rho_\mathrm{S}(t_0)
  \label{rhoSLeff}
\eeq
where ${\cal W}(z)$ is the nonMarkovian effective dot Liouvillian. This we write as
\beq
   {\cal W} (z) = {\cal L}_\mathrm{S} + \Sigma(z),
   \label{WLSig}
\eeq
with ${\cal L}_\mathrm{S}$ describing the free evolution of the system and $\Sigma(z)$ the self-energy or ``memory kernel'' arising from coupling with the leads. 

In the perturbative approach pursued here, the memory kernel is calculated as the series $\Sigma(z) = \sum_{n}\Sigma^{(n)}(z)$ where  $n$ corresponds to the number of interaction Liouvillians ${\cal L}_V$ incorporated in that term.  Tunnelling is governed by the rates
\beq
  \Gamma_{\xi_1 \alpha_1}^{ m_2 m_1} (\omega)
  \equiv 2 \pi \sum_{k_1}
    t_{\bar{1}m_2} t_{1 m_1}
    \delta(\omega-\omega_{k \alpha})
    \label{rates}
    ,
\eeq
the diagonal elements of which are the familiar Fermi golden rule rates
\beq
  \Gamma_{\xi_1 \alpha_1}^{ m_1 m_1} (\omega)
  \equiv 2 \pi \sum_{k_1}
    |t_{1 m_1}|^2
    \delta(\omega-\omega_{k \alpha})
    .
\eeq
In these terms, the expansion of $\Sigma(z)$ is seen as an expansion in small parameter $\widetilde{\Gamma}=\Gamma/k_B T$, such that $\Sigma^{(n)}(z)$ is order $(\widetilde{\Gamma})^{n/2}$.  In the current work, we expand up to fourth order in the coupling Hamiltonian (second order in $\widetilde{\Gamma}$), such that
\beq
  \Sigma(z) \approx \Sigma^{(2)}(z)+ \Sigma^{(4)}(z)
  .
\eeq
The first term describes sequential tunnelling and the second cotunneling.

Details of the calculation of the memory kernel terms are given in Appendix \ref{appDERIV}.
Assuming a constant tunnelling density of states $\Gamma(\omega) = \Gamma$, the sequential term reads
\beq
 \Sigma^{(2)}(z) &=&  
  J^{p_2}_{2 m_2} \frac{
  - p_2 }{z-i \xi_2 (\omega_2 + \mu_2)-{\cal L}_\mathrm{S}} J^{p_1}_{1 m_1}
  \nonumber\\
  && \times t_{2 m_2} t_{1 m_1} \gamma_{21}^{p_2 p_1}
  ,
\eeq
where $\gamma_{21}^{p_2p_1} = \ew{A_2^{p_2}A_1^{p_1}}_{\mathrm{eq.}}$ is an equilibrium reservoir correlation function which evaluates as
\beq
   \gamma_{21}^{p_2p_1} 
  =\delta_{2\overline{1}} p_1 f(-\xi_1 p_1 \omega_1)
  \label{gamma}
  .
\eeq
We then obtain
\beq
 \Sigma^{(2)}(z) &=&  
  - p_1 p_2  
  J^{p_2}_{\bar{1} m_2} \opp{\phi_a}{\phi_a} J^{p_1}_{1 m_1}
  \nonumber\\
  &&
  \times
   \Gamma_{\xi_1 \alpha_1}^{ m_2 m_1}
   I_a^{(2)}(z;\xi_1,p_1,\mu_1)
   ,
\eeq
with the regularised integral
\beq
  I_a^{(2)}(z=O^+ - i \epsilon) &=&  
  \frac{1}{2} f( p_1(\Delta_a + \xi_1\mu_{\alpha_1}-\epsilon))
  \nonumber\\
  &&
  +\frac{i p_1}{2\pi}\phi(p_1(\Delta_a + \xi_1\mu_{\alpha_1}-\epsilon))
  \nonumber
  ,
\eeq
in terms of the function
\beq
  \phi(\lambda) &=& \mathrm{Re} \Psi\rb{\frac{1}{2}+i\frac{\lambda}{2\pi k_BT}} - \ln \frac{D}{2\pi k_BT}
  \label{phidef}
\eeq
with $\Psi$ the digamma function.

The cotunneling term has two contributions: ``direct'' and ``exchange'', such that $ \Sigma^{(4)}(z)  = \Sigma^{(4D)}(z)  + \Sigma^{(4X)}(z) $.  The direct part is given by
\beq
  \Sigma^{(4D)}(z) &=& 
  p_4 p_1 
  J^{p_4}_{\bar{1} m_4} \kett{\phi_a} 
  \braa{\phi_a}
  J^{p_3}_{\bar{2} m_3} 
  \kett{\phi_{a'}} 
  \nonumber\\&&
  \times
  \braa{\phi_{a'}}
  J^{p_2}_{2 m_2}  \kett{\phi_{a''}} \braa{\phi_{a''}}
  J^{p_1}_{1 m_1}
  \nonumber\\
  &&\times
  \Gamma_{1}^{m_4 m_1}\Gamma_{2}^{m_3 m_2}
   \nonumber\\
  &&\times
  I^{D}_{aa'a''}(z;\xi_1,\xi_2,p_1,p_2,\mu_1,\mu_2)
  \label{SDmain}
  ,
\eeq
and the exchange
\beq
  \Sigma^{(4X)}(z)
  &=& 
  - p_4 p_1 
  J^{p_4}_{\bar{2} m_4}\kett{\phi_a} 
  \braa{\phi_a}
  J^{p_3}_{\bar{1} m_3}
   \kett{\phi_{a'}}
  \nonumber\\&&
  \times
  \braa{\phi_{a'}}
  J^{p_2}_{2 m_2} 
   \kett{\phi_{a''}} \braa{\phi_{a''}}
  J^{p_1}_{1 m_1}
  \nonumber\\
  &&\times
  \Gamma_{1}^{m_3 m_1}\Gamma_{2}^{m_4 m_2}
   \nonumber\\
  &&\times
  I^{X}_{aa'a''}(z ;\xi_1,\xi_2,p_1,p_2,\mu_1,\mu_2)
  \label{SXmain}
  .
\eeq
These fourth-order integrals, $I^D$ and $I^X$, are discussed in the Appendix.

\section{Full Counting Statistics}
The density matrix of \eq{rhoSLeff} is the Laplace-transform of the solution to the nonMarkovian master equation \cite{FN_nonMark}
\beq
   \dot{\rho}_\mathrm{S}(t) = \int_0^t dt' {\cal W} (t-t')\rho(t')
   \label{NMQME}
   .
\eeq
In contrast to e.g. Ref.~\cite{lei08}, we make no further approximations at this point --- in particular, we do not make Markov approximation --- but rather seek to calculate the FCS of \eq{NMQME} as it stands.
This requires that we introduce counting fields $\chi_\alpha$ in the appropriate places in the Liouvillian, a process which yields the ``$\chi$-resolved'' Liouvillian  ${\cal W}(\chi,z)$ \cite{bag03}.

In the current Liouville scheme, this can be achieved by replacing each contraction $\gamma_{21}^{p_2 p_1}$ in the memory kernel $\Sigma(z)$ by the counting-field-dependent analogue $\gamma_{21}^{p_2p_1}(\chi)$, which we define as
\beq
   \gamma_{21}^{p_2p_1}(\chi)=\gamma_{21}^{p_2p_1}
   \exp
   \left[
     i s_{\alpha_1} \xi_1 \rb{\frac{p_1-p_2}{2}} \chi_{\alpha_1}
   \right]
   \label{gamchi}
   ,
\eeq
with $s_\alpha=\pm 1$ a factor given by the sign-convention for current flow in lead $\alpha$. In the following we will only count in a single lead, for which we choose $s_{\alpha}=1$.
To see that $\gamma_{21}^{p_2p_1}(\chi)$ adds a counting field at the correct points, consider, for example, the case with $-\xi_2 = \xi_1 = +$. Then, for $-p_2=p_1=+$, the contraction $ \gamma_{21}^{p_2p_1}$ is proportional to trace of $ a_{k\alpha}^\dag \rho a_{k\alpha} $, a state with one more electron in lead $\alpha$ than $\rho$ itself. On the other hand, for $p_2=p_1=+$, we require the trace of $ a_{k\alpha}^\dag a_{k\alpha} \rho $, a state with the same number of electrons as $\rho$. The required counting-field factors are $e^{i s_\alpha \chi_\alpha}$ and $e^0=1$, respectively, as given by \eq{gamchi}.

Once in possession of the ``$\chi$-resolved'' Liouvillian, the cumulant generating function (CGF) ${\cal F}(\chi) = - t z_0(\chi)$ is obtained from the solution $z_0(\chi)$ of the equation
\beq
  z_0 - \lambda_0(\chi;z_0) = 0
  \label{z0lam}
  ,
\eeq
where $\lambda_0(\chi;z_0)$ is the eigenvalue of ${\cal W}(\chi;z)$ that develops adiabatically from zero as $\chi$ is increased from zero \cite{jau05}.
In the Markovian case, $\lambda_0$ is independent of $z$, and the CGF is simply ${\cal F}(\chi) = - t \lambda_0(\chi)$.  The nonMarkovian case is less straightforward, however.
We follow here the approach of Ref.~\cite{fli08}, which uses \eq{z0lam} to derive expressions for the cumulants themselves, by-passing an explicit evaluation of the CGF itself.  This approach can deliver the cumulants up to, in principle, arbitrary order ($> 20$ in Ref.~\cite{fli04}) and without any further approximation.

We first define
\beq
  {\cal J}(\chi,\epsilon) = {\cal W}(\chi,z=0^+-i\epsilon)-{\cal W}(\chi=0,z=0^+)
  ,
\eeq
along with the derivatives
\beq
  {\cal J}'= \left. \partial_{\chi} {\cal J}\right|_{\chi,\epsilon \to 0}
  ;\quad\quad
   \dot{\cal J} = \left. \partial_\epsilon {\cal J}\right|_{\chi,\epsilon \to 0}
   ,
\eeq
and analogously for higher-orders. We define the left and right null-vectors of ${\cal W}(0,0^+)$ via
\beq
  {\cal W}(0,0^+) \kett{\psi_0} = \braa{\psi_0} {\cal W}(0,0^+) =0
  ,
\eeq
which we assume to be unique. The vector $\kett{\psi_0}$ corresponds to the stationary density matrix of the system, and multiplication with $\braa{\psi_0}$, corresponds to taking the trace over system states\cite{jau05}.  We define the stationary state ``expectation value'' $\eww{\bullet} = \eww{\psi_0 | \bullet |\psi_0}$, and the projectors ${\cal P} = \opp{\psi_0}{\psi_0}$ and ${\cal Q} = \mathbbm{1}-{\cal P}$. Finally, we require the pseudo-inverse
\beq
  {\cal R}(\epsilon) =  {\cal Q} \frac{1}{i \epsilon + L(0,0^+ -i\epsilon)} {\cal Q}
  \label{pseudoR}
  .
\eeq

From Refs. \cite{fli08,flindtthesis}, the first three current cumulants are
\beq
  \eww{I^1} &=& \eww{I^1}_m
  \\
  \eww{I^2} &=& \eww{I^2}_m 
  +2 \eww{I^1}_m 
    \eww{
      \dot{{\cal J}}' - {\cal J}' {\cal R} \dot{{\cal J}}
    }
  \\
  \eww{I^3} &=& \eww{I^3}_m 
  - \frac{3\eww{I^2}}{2 \eww{I^1}_m} \rb{\eww{I^2}_m - \eww{I^2}}
  \nonumber\\
  &&
  - 3 i \eww{I^1}_m
  \eww{
    \dot{{\cal J}}''- 2 \dot{{\cal J}}' {\cal R} {\cal J}' - {\cal J}'' {\cal R} \dot{{\cal J}}
  }
  \nonumber\\
  &&
  + 6 i \eww{I^1}_m
  \eww{
   {\cal J}' {\cal R} \rb{{\cal R} \dot{{\cal J}} {\cal P} {\cal J}' + \dot{{\cal J}}'}
  }
  \nonumber\\
  &&
  - 6 i \eww{I^1}_m
  \eww{
   {\cal J}' {\cal R} \rb{ {\cal J}' {\cal R} \dot{{\cal J}} + \dot{{\cal J}} {\cal R} {\cal J}' }
  }
  \nonumber\\
  &&
  +3 i \rb{\eww{I^1}_m}^2
  \eww{
    \ddot{{\cal J}}'
    + 2 {\cal J}' {\cal R} \dot{{\cal J}} {\cal R} \dot{{\cal J}}
  }
  \nonumber\\
  &&
  -3 i \rb{\eww{I^1}_m}^2
  \eww{
    {\cal J}' {\cal R} \ddot{{\cal J}} + 2 \dot{{\cal J}}' {\cal R} \dot{{\cal J}} 
  }
  ,
\eeq
where
\beq
  i \eww{I^1}_\mathrm{m} &=& \eww{{\cal J}'}
  \\
  i^2 \eww{I^2}_\mathrm{m} &=& \eww{{\cal J}''-2 {\cal J}' {\cal R} {\cal J}'}
  \\
  i^3 \eww{I^3}_\mathrm{m} &=& 
  \eww{
    {\cal J}''' - 3 {\cal J}''{\cal R} {\cal J}' - 3{\cal J}' {\cal R} {\cal J}'' 
  }
   \nonumber\\
   &&
  - 6 
  \eww{
    {\cal J}' {\cal R} ({\cal R} {\cal J}' {\cal P}-{\cal J} ' {\cal R}){\cal J}'
  }
  ,
\eeq
are the cumulants in the Markov approximation.  In these expressions, it is understood that the pseudo-inverse is evaluated at $\epsilon=0$.
Although it is only practicable to explicitly write down the cumulants up to third order, the high-order cumulants can be obtained recursively \cite{fli08}.

Reference~\cite{bra06} took a different approach to calculating the cumulants.  There it was assumed that ${\cal W}$ is known to a given order in some small parameter, and the CGF is then calculated to the same order. For problems such as that considered here, this means that the CGF, and hence all the cumulants, are calculated rigorously up to order $(\widetilde{\Gamma})^2$. This is to be contrasted with the above cumulants which, if expanded, have contributions at all orders in $(\widetilde{\Gamma})$.
Whilst the method of Ref.~\cite{bra06} may seem more consistent, there are two good reasons why the approach described here might be preferable. Firstly, from Ref.~\cite{gur96} we know that in the infinite bias limit, the effective Liouvillian is given exactly by ${\cal L}_0$ plus the rate part of $\Sigma^{(2)}$.  In order to recover the FCS correctly in this limit then, no further approximations should be made when calculating the cumulants \cite{FNDQD}.
Secondly, Ref.~\cite{lei08} makes the point that, in certain circumstances, by treating incoming and outgoing processes unequally, a strict order-by-order approach can lead to unphysical results for the current.  In the next section we shall compare these two methods directly.

Before doing so, we note that comparing the foregoing expressions for the current and the shotnoise with those of, e.g., Ref.~\cite{thi05}, we can identify the derivatives of ${\cal J}$ with respect to $\chi$ and $z$ with the various ``current super-operator blocks'' of the real-time diagrammatic approach.
For example, differentiating  ${\cal J}$ once with respect to $i\chi$ and setting $\chi\to 0$ yields a super-operator like $\cal W$, but with an additional forefactor $\xi_1 (p_1-p_2)/2$.  Under the summation, the $p_2$ contribution cancels, leaving a forefactor $\xi_1p_1/2$.  This forefactor is the same as arises in replacing a single tunnel vertex with a current vertex in the self-energy, which is the recipe for obtaining the current super-operator block used to calculate the stationary current in the diagrammatic approach.
Using a very different approach, we thus reproduce the real-time current and shotnoise expressions. The advantage of the present method is that it is now easy to obtain to the higher cumulants, whereas with the diagrammatic approach, this requires some effort.

\section{Transport through a single quantum dot}

We model the transport through a single Zeeman-split level with the 
Anderson Hamiltonian \cite{and61}
\beq
  H = 
   \sum_\sigma \epsilon_\sigma d_\sigma^\dagger d_\sigma
  + U n_\uparrow n_\downarrow
  + H_\mathrm{res}
  + V,
\eeq
where $\epsilon_\sigma$ is the energy of a spin-$\sigma$ electron in the dot and $U$ is the interaction energy.  The reservoir and interaction Hamiltonians are as \eq{Hres} and \eq{V1}, with index $\alpha$ including both lead ($=L,R$) and spin index.
In the limit of large level-splitting we can address one and only one Zeeman level.  We then recover the SRL model, the transport properties of which can be obtained exactly from scattering theory \cite{but92,bla00}.  The SRL thus provides a useful benchmark against which to compare our approximate methods.

We will discuss results obtained in several different approximate schemes.  We denote as ``O(4)'' the results obtained by calculating the cumulants as outlined in the previous section with the fourth-order effective Liouvillian containing both sequential and cotunneling terms.  The second-order ``O(2)'' solution is obtained in the same way, but with sequential terms only.  We also consider a scheme in which we expand the cumulants  and truncate at fourth order.   In this way we recover the FCS results of Braggio {\it et al} \cite{bra06} and the shotnoise results of Thielmann {\it et al} \cite{thi05}.  This approach we label as ``O(4) trunc''.
Finally, we compare with results in the Markovian approximation, which we label with ``Mark.''.

We calculated results with and without the level-renormalisation parts of the fourth-order self-energy (integrals $I^{D0}$ and $I^{X0}$, and $I^{D2}$ and $I^{X2}$  of the Appendix). Their contribution was found to be negligible in all cases studied here. In the results presented below, these parts of the self-energy have been neglected.  This reduces considerably the computational effort involved since the double-principal-part integrals ($I^{D0}$ and $I^{X0}$) must be evaluated numerically.

\subsection{Single resonant level}

The calculation of the first three cumulants in the scattering approach is discussed in Appendix \ref{appSCAT}.  In the infinite bias limit, we have \cite{jon96}
\beq
  \ew{I} &=& \frac{\Gamma_L \Gamma_R}{\Gamma}
  ;\quad
  S = \ew{I} \frac{\Gamma_L^2 + \Gamma_R^2}{\Gamma^2};
  \nonumber\\
  S^{(3)} &=& \frac{\ew{I}}{\Gamma^4}
  \rb{
    \Gamma_L^4 - 2 \Gamma_L^3 \Gamma_R + 6 \Gamma_L^2 \Gamma_R^2  -2 \Gamma_L \Gamma_R^3 + \Gamma_R^4
  }
  ,
  \nonumber
\eeq
or $\ew{I}=\Gamma_L/2$,  $S=\Gamma_L/4$ and  $S^{(3)}=\Gamma_L/8$ for symmetric coupling, $\Gamma_L=\Gamma_R$.

\begin{figure}[tb]
  \psfrag{GkBTeq1}{$\widetilde{\Gamma}=1$}
  \psfrag{GkBTeq12}{$\widetilde{\Gamma}=1/2$}
  \psfrag{GkBTeq14}{$\widetilde{\Gamma}=1/4$}
  \psfrag{eVkBT}{$e V/k_B T$}
  \psfrag{IkBT}{$\widetilde{I}$}
  \psfrag{SkBT}{$\widetilde{S}$}
  \begin{center}
  \epsfig{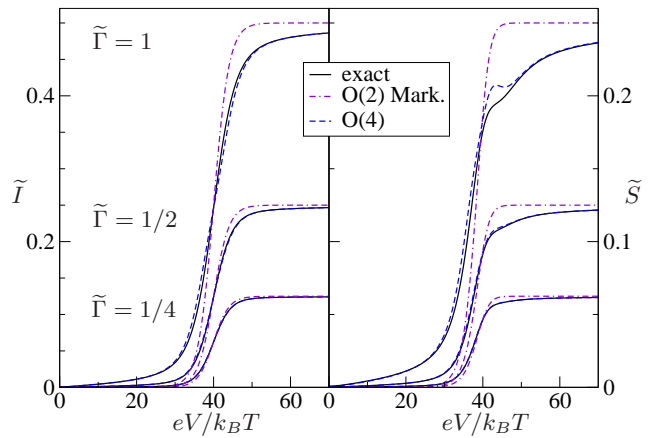}
  \caption{
    Stationary current $\widetilde{I}=\ew{I}/k_BT$ (left) and zero-frequency shotnoise $\widetilde{S}=S/k_BT$ (right) as a function of applied bias $eV$ for the single resonant level model with level located at $\varepsilon=20 k_B T$, chemical potentials $\mu_L=-\mu_R =  e V/2$, and bandwidth $D=10^3 k_B T$. Results are shown for three different couplings: $\widetilde{\Gamma}=\Gamma_L/k_B T=\Gamma_R/k_B T= \frac{1}{4},\frac{1}{2},1 $ and three calculational schemes: exact, full 4th-order (O(4)), and 2nd-order Markovian (O(2) Mark.).  Whereas the O(2) Markovian results show obvious deviations from the exact results for these couplings, the O(4) solution gives good agreement except around the top of the shotnoise step for $\widetilde{\Gamma}=1$.
    \label{SRLSnIfig}
 }
  \end{center}
\end{figure}
Figure \ref{SRLSnIfig} shows the current and shotnoise as a function of applied bias for different values of the coupling $\widetilde{\Gamma}=\Gamma_L/k_B T=\Gamma_R/k_B T$.  A step occurs at $e V \approx 2\varepsilon$ ($ = 40 k_B T$ here) as the level enters the transport window.
For $\widetilde{\Gamma}\lesssim \frac{1}{4}$ agreement between the our O(4) fourth-order calculation and the exact result is excellent across the whole bias range.  For greater couplings, $\widetilde{\Gamma} \gtrsim 1/2$,  deviation from the exact solution is seen in the shotnoise around the top of the step which signals the start of the break down of our approach.
The difference between the second-order Markovian and the exact solution is stark.  In the CB regime ($e V \lesssim 30$ here) the sequential current is almost totally suppressed, but the cotunneling current is still considerable. The O(2) Mark. solution also shows significant error in the high-bias ($e V \gtrsim 2\varepsilon$) regime, which arises largely from the Markovian approximation.

\begin{figure}[tb]
  \psfrag{GkBTeq1}{$\widetilde{\Gamma}=1$}
  \psfrag{GkBTeq12}{$\widetilde{\Gamma}=1/2$}
  \psfrag{GkBTeq14}{$\widetilde{\Gamma}=1/4$}
  \psfrag{eVkBT}{$e V/k_B T$}
  \psfrag{S3kBT}{$\widetilde{S}^{(3)}$}
  \begin{center}
  \epsfig{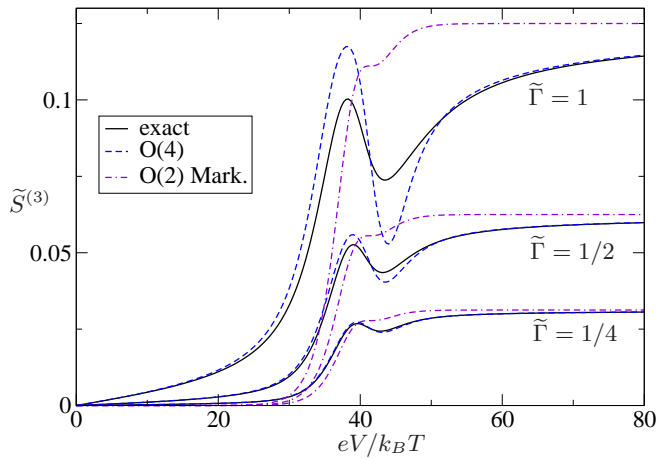}
  \caption{
    Zero-frequency skewness  $\widetilde{S}^{(3)}=S^{(3)}/k_BT$ of SRL with the same parameters as in \fig{SRLSnIfig}. For a given coupling, agreement with the exact solution is worse than for shotnoise, but still good at low couplings (e.g. $\widetilde{\Gamma}=1/4$).
    \label{SRLS3fig}
 }
  \end{center}
\end{figure}

\begin{figure}[tb]
  \psfrag{eVkBT}{$e V/k_B T$}
  \psfrag{SkBT}{$\widetilde{S}$}
  \psfrag{F}{$F$}
  \psfrag{Geq12}{$\widetilde{\Gamma} = 1/2$}
  \begin{center}
  \epsfig{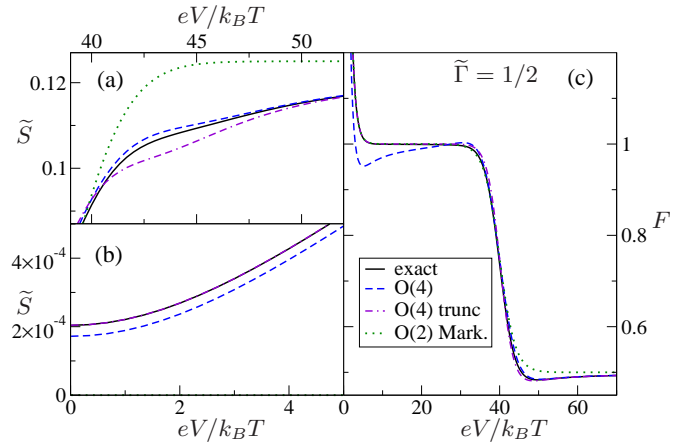}
  \caption{
    Zero-frequency shotnoise (left panels) and Fano factor $F=S/\ew{I}$ (right panel) of SRL as a function of applied bias $eV$.  The tunnel rate is fixed at $\widetilde{\Gamma} =1/2$ here, but otherwise the parameters are as \fig{SRLSnIfig}. 
    In addition to the approximation schemes discussed in \fig{SRLSnIfig}, results are also shown here from a rigorous expansion of the cumulants to 4th order (O(4) trunc).  
    Around the current step (panel (a)), the full O(4) solution describes the behaviour better than O(4) trunc. 
    However, in the low bias, Coulomb blockade regime (panel (b)), it is the O(4) trunc solution that matches the exact solution better.
    Panel (c) illustrates the dangers of considering the Fano factor alone: although O(4) trunc reproduces the exact Fano factor extremely well in the Coulomb blockade regime, so does the O(2) Markovian solution, which we know gives the current and shotnoise individually extremely poorly.
    \label{SRLSFfig}
 }
  \end{center}
\end{figure}

\begin{figure}[tb]
  \psfrag{GkBT}{$\Gamma_L/k_B T$}
  \psfrag{eVkBT}{$e V/k_B T$}  
  \psfrag{S3kBT}{$\widetilde{S}^{(3)}$}
  \psfrag{F}{$F^{(3)}$}
   \psfrag{Geq12}{$\widetilde{\Gamma} = 1/2$}
  \begin{center}
  \epsfig{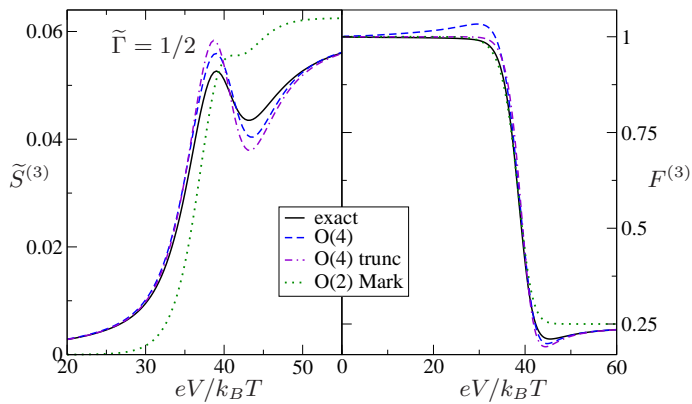}
  \caption{
    Zero-frequency skewness (left) and associated Fano factor $F^{(3)}=S^{(3)}/\ew{I}$ (right) of the SRL model as a function of applied bias $eV$.  Same parameters and labels as \fig{SRLSFfig}. Once again, the O(4) solution performs better in the onset region.
    \label{SRLS3F3fig}
 }
  \end{center}
\end{figure}

Figure \ref{SRLS3fig} plots the skewness which show a pronounced undulation at the onset.
The O(2) Markovian solution provides only the coarsest description of this behaviour, whereas it is reproduced by the O(4) solution.  For $\widetilde{\Gamma} \lesssim \frac{1}{4}$, the quantitative agreement with the exact result is good.  Nevertheless, for a given coupling, the error is larger for skewness than for the noise. This is not too surprising, since we expect higher-order correlators to be sensitive to higher-order processes, which are of course neglected here.  The implication is that for a given coupling, only cumulants up to a certain order can reliably be calculated from any approximate effective Liouvillian \cite{FNinf}.
Figures \ref{SRLSFfig} and \ref{SRLS3F3fig} compare the O(4) and O(4)~trunc results at $\widetilde{\Gamma}=1/2$.  We choose this value to highlight the differences between the two solutions which, for smaller couplings, are negligible.
Near the top of the step (\fig{SRLSFfig}a and \fig{SRLS3F3fig}a), the O(4)  and O(4) trunc solutions are noticeably different, with our O(4) results significantly closer to the exact result.  
Deep in the CB regime (\fig{SRLSFfig}b), the two approximate solutions differ once more, but this time the trunc solution is more accurate. 
The difference is small (a difference in $\widetilde{S}$ of $2\times 10^{-5}$ near zero bias); it plays, however, a disproportionate role in determining the Fano factors in the CB regime as \fig{SRLSFfig}c and \fig{SRLS3F3fig}b show.  
From the exact solution, we know that noise Fano factor diverges at low bias (fluctuation dissipation theorem), whereas the skewness Fano factor 
$
  F^{(3)} \equiv S^{(3)}/\ew{I}
$ tends to unity.
In this regime, both Fano factors are reproduced better by trunc solution than by solution O(4). It should be borne in mind that, in obtaining the Fano factors in this region, one is dividing one very small quantity by another and thus even small absolute errors can effect Fano factors quite dramatically. As a warning not to take the finer details of the Fano factor too seriously, we observe that the O(2) Mark.~solution in the CB regime reproduces the exact Fano factors better than any of the fourth-order results. This is deceptive since the individual current and shotnoise obtained with this method are vastly different from the exact results.

From this comparison we obtain a degree of confidence in the cumulants of up to at least third order for $\widetilde{\Gamma}=1/4$.  Our O(4) solution appears to perform better in the region where levels are crossing the chemical potentials, whereas the O(4) trunc solution gives better results in the CB regime. In this regime, the transport properties are effectively Markovian, which can be demonstrated by comparing fourth-order solutions with and without the Markov approximation..

\subsection{Anderson Model}

\begin{figure}[tb]
  \psfrag{eVkBT}{$e V/k_B T$}
  \psfrag{IkBT}{$\widetilde{I}$}
  \psfrag{SkBT}{$\widetilde{S}$}
  \psfrag{F}{$F$}
  \psfrag{Geq14}{$\widetilde{\Gamma} = 1/4$}
  \begin{center}
  \epsfig{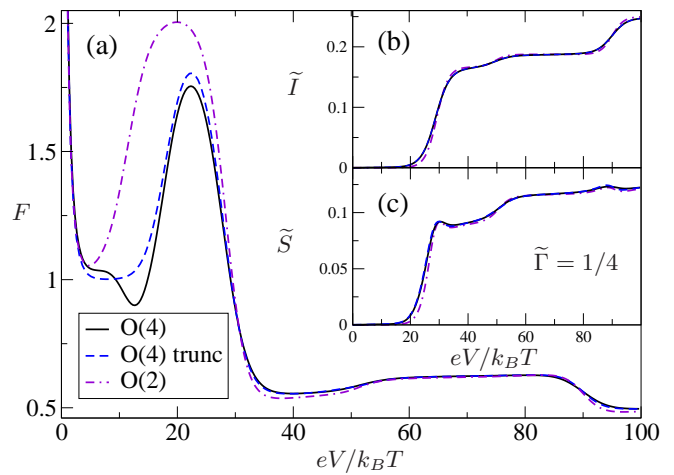}
  \caption{
    Current (b), noise (c) and Fano factor (a) for the Anderson model as a function of applied bias $eV$.
    Parameters were  chosen as in \cite{thi05}: $\Gamma_L=\Gamma_R= \frac{1}{4} k_B T$,
    $\mu_L=-\mu_R = \frac{1}{2} e V$, $\epsilon_\uparrow=-15 k_B T$, $\epsilon_\downarrow=5 k_B T$, $U=40 k_B T$ and bandwidth $D=10^3 k_B T$.
    A prominent peak is observed in the Fano factor around a bias such that the transport window includes the upper level whilst the lower level is still included.  Cotunneling reduces the size of the peak.
    \label{ANDF2fig}
 }
  \end{center}
\end{figure}

\begin{figure}[tb]
  \psfrag{eVkBT}{$e V/k_B T$}
  \psfrag{F3kBT}{$F^{(3)}$}
  \psfrag{S3kBT}{$\widetilde{S}^{(3)}$}
  \psfrag{Geq14}{$\widetilde{\Gamma} = 1/4$}
  \begin{center}
  \epsfig{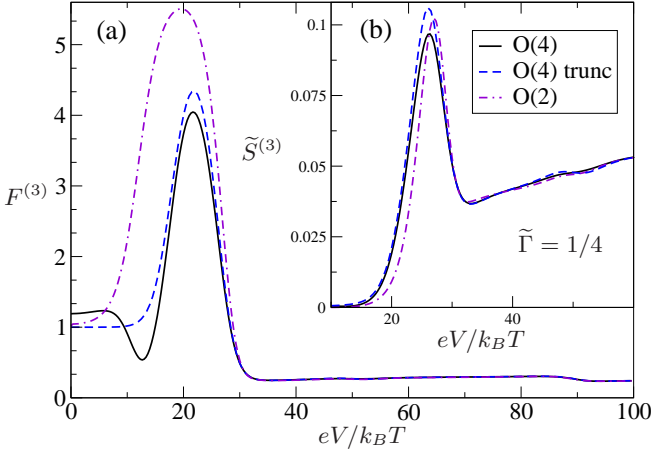}
  \caption{
    Skewness (b) and skewness Fano factor $F^{(3)}$ (a) for the Anderson model with parameters and labelling as \fig{ANDF2fig}. Not only the skewness Fano factor, but the skewness itself show a large peak as the top level enters transport window.
    \label{ANDF3fig}
 }
  \end{center}
\end{figure}

The current, shotnoise and Fano factor of the Anderson model with cotunneling were investigated in Ref.~\cite{thi05}, and, as Fig.~\ref{ANDF2fig} shows, the present calculation broadly reproduces these results.  The situation in which the lower dot level lies below the transport window is of particular interest.  As observed in Ref.~\cite{thi05}, increasing the applied bias results in a large peak in the Fano factor around the point where the upper dot level enters the transport window.
The peak exists in the sequential tunnel limit, but its height, width, and location are markedly altered by inelastic cotunneling processes. No peak occurs in the shotnoise itself; only in the Fano factor is this feature visible.
Figure \ref{ANDF3fig} shows our results for the skewness in this situation.  It is immediately clear that the skewness Fano factor also shows a peak, and that this is even more pronounced than that of the noise.  Furthermore, the skewness itself exhibits a sharp peak, as inset \fig{ANDF3fig}b shows.  As for the noise Fano factor, the presence of cotunneling significantly reduces the height and overall area of the peak.

This superPoissonian behaviour indicates a significantly bunched electron flow, which can nicely be explained with the dynamical channel blockade model of \cite{bel04}, in which a single level (here, the lower) is but weakly coupled to the collector.  In the simple sequential picture of Ref.~\cite{bel04}, the shotnoise and skewness Fano factors are predicted to be $F_2 = (1+p)/(1-p)$ and $F_3 =(1+4p+p^2)/(1-p)^2$, where $1/p$ is parameter corresponding to the number of ways in which the dot can be filled.  With $p=1/3$ (corresponding to three ways of filling the dot: from the left and right into the lower level, and from the left only into the upper level), we obtain $F_2 = 2$ and $F_3 =11/2$, which are almost exactly the values obtained by our sequential O(2) results at the tops of the peaks.  Cotunneling reduces the heights of peaks, and good agreement with the dynamical channel blockade model can be obtained with the choice $p=0.272$.

Both noise and skewness figures show results for O(4) and O(4)~trunc solutions.
At high bias, these solutions agree closely with one another and both predict the same position and widths of the Fano factor peaks.  However, the heights of the peaks are given differently in the two approaches, with the O(4) trunc peaks being somewhat higher. Differences between the two solutions are most pronounced at low bias, when the Coulomb blockade is in effect.  For example, the O(4) solution predicts a small subPoissonan dip before the superPoissonian peak, which is absent in the O(4)~trunc results.  From our studies of the SRL model, we expect the O(4)~trunc prediction to be more accurate in this regime, and this is borne out by the fact that the skewness Fano factor in the O(4) calculation does not tend to unity with decreasing bias.
Conversely, based on the SRL results, we expect the height of the Fano factor peaks to be better described by the O(4) solution, i.e. the lower of the two values.

\section{Conclusions \label{concs}}

We have described a method for calculating the counting statistics of an arbitrary mesoscopic system that takes into account both sequential tunnelling and cotunneling of electrons.  This is achieved by performing a perturbative expansion of the von Neumann equation in Liouville-Laplace space and adding counting fields to the reservoir contractions. Current cumulants are then obtained without further approximation with the pseudo-inverse approach.
In principle, cumulants of arbitrary order can be calculated in this fashion.  However, as we expect higher-order cumulants to be sensitive to higher-order tunnel processes, there will be an inevitable reduction in accuracy as the order of the cumulant increases.
Use of the pseudo-inverse means that this method is applicable to systems of large size, unlike methods which explicitly require the eigenvalue $\lambda_0(\chi,z)$ of the effective Liouvillian.

We have studied here transport through a single quantum dot with both sequential and cotunneling contributions.  Of particular interest is the single-resonant level model as it provides an exact case against which we can compare.  Good agreement with the exact results was found for both shotnoise and skewness up to ratio between coupling rate and reservoir temperature of $\widetilde{\Gamma} \lesssim 1/4$.  Moreover, we also compared our results with those obtained from a rigorous expansion of the cumulants up to fourth order in the tunnel coupling.  From this comparison we conclude that for intermediate biases, and in particular when system levels lie near the chemical potentials of the leads, the approach described here gives better results. In the Coulomb blockade regime, however, the rigorous fourth-order approach is better, but here the dynamics are effectively Markovian. In general, the difference between these two different fourth-order approximations is far less than that between fourth and second order approximations, and decreases with decreasing system-reservoir coupling.

Future work includes the study of transport models with internal quantum degrees of freedom, such as the double quantum dot, to understand how the interplay of cotunneling and internal dynamics effects counting statistics.

\appendix

\section{Derivation of effective system Liouvillian \label{appDERIV}}

Following Refs.~\cite{lei08,kor07,sch08}, we start by defining the system operators
\beq
  g_{k\alpha} &=& 
  \sum_{m}  t_{k \alpha m} j_{m}
  \label{gtj}
  ,
\eeq
such that the interaction Hamiltonian of \eq{V1} can be written
$V = \xi_1 a_1 g_1$.
Correspondingly, in Liouville space we have
\beq
  {\cal L}_V =-i \xi_1 \sum_p p \sigma^p A^p_1 G^p_1
  \label{LVdef}
\eeq
with $A$ as before, and $G$ defined via 
\beq
  G^p_1 O = 
  \sigma^p 
  \left\{
   \begin{array}{c c}
      g_1 O, &p =+ \\
      - O g_1, &p =-
    \end{array}
  \right.
  \label{Gdef}
  .
\eeq
The object $\sigma^p$ is a dot-space superoperator with matrix elements \cite{sch08}
\beq
  \rb{\sigma^p}_{ss',\bar{s}\bar{s}'} = 
  \delta_{s\bar{s}}\delta_{s'\bar{s}'} 
  \left\{
   \begin{array}{c c}
      1,\quad N_s - N_{s'}=~\mathrm{even} \\
      p,\quad N_s-N_{s'}=~\mathrm{odd}
    \end{array}
  \right.
  ,
\eeq
where, $N_s$ is the number of electrons in state $s$.
Note that $G_1^p =  p \sigma^p t_{1m} J_{1m}^p$.

The reduced density matrix of the dot is given by tracing out the electron reservoirs
\beq
  \rho_\mathrm{S}(z) 
  = 
  \mathrm{Tr}_\mathrm{R} \left\{\frac{}{} \rho(z) \right\}
  = 
  \mathrm{Tr}_\mathrm{R}  
  \left\{\frac{}{}
    \frac{1}{z-{\cal L}}  \rho(0)
  \right\}
  .
\eeq
This we expand in powers of ${\cal L}_V$ to obtain
\beq
  \rho_\mathrm{S}(z) 
  =  
  \mathrm{Tr}_\mathrm{R}  
  \left\{
    \rb{
      \Omega_0(z)
      +
      \Omega_0(z){\cal L}_V\Omega_0(z)
      +\ldots
    }
    \rho(0)
  \right\}
  \label{rhoD_exp}
\eeq
with free propagator
$
  \Omega_0(z) = \left[z-{\cal L}_\mathrm{res}-{\cal L}_\mathrm{S}\right]^{-1}$.
With substitution of \eq{LVdef}, a typical term of the expansion  \eq{rhoD_exp} reads
\beq 
  (-i)^n
  \rb{\prod_{l=1}^n \xi_l p_l}
  \mathrm{Tr}_\mathrm{R}  
  \left\{\frac{}{}
    \Omega_0(z)
    \sigma^{p_n}  A^{p_n}_n G^{p_n}_n
    \ldots
    \right.
    \nonumber\\
    \ldots
    \left.
    \sigma^{p_1} A^{p_1}_1 G^{p_1}_1
    \Omega_0(z)
    \rho(t_0)
    \frac{}{}
  \right\}
  .
\eeq
Evaluating the action of the $\sigma^p$ superoperators we obtain a factor $\prod_{l}^{\mathrm{odd}} p_l$ and, as the $G$ operators also contain $\sigma$, they evaluate at different positions in the chain as
\beq
  G^{p_{l}}_{l}O 
  &=& 
  (p_l)^{l+1}
  \left\{
   \begin{array}{c}
      g_l O \\
      O g_l
    \end{array}
  \right.
  .
\eeq
Our typical term then looks like
\beq 
  (-i)^n
  \rb{\prod_{l}^\mathrm{all} \xi_l p_l}
  \rb{\prod_{l'}^{\mathrm{odd}} p_{l'}}
  ~~~~~~~~~~~~~~~~~~~~~~~~~~~
  \nonumber\\
  \times
  \mathrm{Tr}_\mathrm{R}  
  \left\{\frac{}{}
    \Omega_0(z)
    A^{p_n}_n G^{p_n}_n 
    \ldots
    A^{p_1}_1 G^{p_1}_1 
    \Omega_0(z)
    \rho(t_0)
  \right\}
  \label{typ_ps}
  ,
\eeq
and the next task is to separate dot and reservoir degrees of freedom.  For this we can use the dot-reservoir superoperator commutation relation
$
  A_1^p G_{1'}^{p'} = - p p'  G_{1'}^{p'} A_1^p
$.  We will also need the following reasons: 
$
  \mathrm{Tr}_\mathrm{R} {\cal L}_\mathrm{res} 
  = 0
$
,
$  
  {\cal L}_\mathrm{res} \rho_\mathrm{res}^\mathrm{eq} =0
$
and
$
  A_1^p  {\cal L}_\mathrm{res} = \rb{  {\cal L}_\mathrm{res} - x_1 }A_1^p 
$
with
\beq
  x_1 &=& -i \xi_1 (\omega_1 + \mu_{\alpha_1})
.
\eeq
The commutation of the dot-operators through the free propagators therefore changes the argument of the propagator:
\beq
  A_1^p \Omega_0(z) =
 \Omega_0(z+x_1) A_1^p
 \label{dOm}
 .
\eeq

Bringing all the $A$ operators to the right of the $G$ operators generates a factor which exactly cancels with the first product in \eq{typ_ps}.  Our term becomes
\beq
  &&
  (-i)^n \rb{\prod_{l'}^{\mathrm{odd}} p_{l'}}
  \mathrm{Tr}_\mathrm{R}  
  \left\{\frac{}{}
    \Omega_\mathrm{S}(z)
    G^{p_n}_n 
    \Omega_\mathrm{S}(z_{n-1})
    G^{p_{n-1}}_{n-1} 
    \ldots
  \right.
  \nonumber\\
  &&
  \left.
    \ldots 
     G^{p_2}_2 
    \Omega_\mathrm{S}(z_1)
    G^{p_1}_1 
    \Omega_\mathrm{S}(z)
    A^{p_n}_n
   \ldots
    A^{p_1}_1
    \rho(0)
  \right\}
  \label{typ_X}
\eeq
with free dot propagator
\beq
  \Omega_\mathrm{S}(z) = \frac{1}{z-{\cal L}_\mathrm{S}}
  ,
\eeq
and 
$
  z_m = z+ \sum_{l=m+1}^{n} x_l ;
  \quad
  1 \le m \le n-1
$.

The the reservoir expectation values,
$
  \mathrm{Tr}_\mathrm{res}  
  \left\{\frac{}{}
    A^{p_n}_n
   \ldots
    A^{p_1}_1
    \rho_\mathrm{res}^\mathrm{eq}
  \right\}
  =
  \ew{A^{p_n}_n
   \ldots
    A^{p_1}_1
    }_\mathrm{eq}
$, can be evaluated with the rules of Wick's theorem in Liouville space, which read \cite{lei08}
\begin{itemize}
  \item decompose $\ew{\ldots}_\mathrm{res}$ into pair-contractions
  \item add minus sign for each interchange of $A$
  \item omit factor $ \rb{\prod_{l'}^{\mathrm{odd}} p_{l'}}$ arising from 
    $\sigma$ super-operators
  \item each pair contraction then contributes a factor
  \beq
   \ew{A_2^{p_2}A_1^{p_1}}=
  \gamma_{21}^{p_2p_1}=\delta_{2\overline{1}} p_1 f_{\alpha_1}(-\xi_1 p_1 \omega_1)
  .
  \eeq
\end{itemize}
With these rules, our typical term becomes
\beq
  (-i)^n 
  \Omega_\mathrm{S}(z_n)
  G^{p_n}_n 
  \Omega_\mathrm{S}(z_{n-1})
  G^{p_{n-1}}_{n-1} 
  \ldots 
  \quad\quad \quad\quad
  \nonumber\\
  \quad\quad\quad
  \ldots
   G^{p_2}_2 
  \Omega_\mathrm{S}(z_1)
  G^{p_1}_1 
  \Omega_\mathrm{S}(z)
  \rho_\mathrm{S}(t_0)
   \quad\quad 
  \nonumber\\
  \quad\quad \quad\quad 
  \times
  \rb{
    \sum_\mathrm{decomps} \rb{-1}^{N_P} \prod \gamma_{ij} 
  }
  \label{typ_Wick}
 ,
\eeq
where the last factor indicates a sum over all pair decompositions with the relevant Wick sign $(-1)^{N_P}$.

Comparison of this expression with \eq{rhoSLeff} and \eq{WLSig} allows us to identify the self-energy as
\beq
  \Sigma(z) &=&  (-i)^n
  \sum_{n}^\mathrm{even}
  \rb{
    \sum_\mathrm{irred.} \rb{-1}^{N_P} \prod \gamma_{ij} 
  }
  \nonumber\\
  &&
  \times
  G^{p_n}_n 
  \Omega_\mathrm{S}(z_{n-1})
  G^{p_{n-1}}_{n-1} 
  \ldots 
   G^{p_2}_2 
  \Omega_\mathrm{S}(z_1)
  G^{p_1}_1 
  \nonumber
  ,
\eeq
where the sum is over {\it irreducible} contractions only.

\subsection{Second order}

At second order, there is only one contraction, and we have 
\beq
  \Sigma^{(2)}(z) &=& G^{p_2}_2 \frac{-1}{z+x_2 -{\cal L}_\mathrm{S}} G^{p_1}_1 \gamma_{21}^{p_2 p_1}
  \nonumber\\
  &=&
  G^{p_2}_{\bar{1}} \frac{- p_1 f_{\alpha_1}(-\xi_1 p_1 \omega_1)}{z+i\xi_1(\omega_1+\mu_{\alpha_1}) -{\cal L}_\mathrm{S}} G^{p_1}_1
  \nonumber\\
  &=&
  - G^{p_2}_{\bar{1}} \opp{\phi_a}{\phi_a}G^{p_1}_1  
  \frac{p_1 f_{\alpha_1}(-\xi_1 p_1 \omega_1)}
    {z+i\xi_1(\omega_1+\mu_{\alpha_1}) +i\Delta_a}
    \nonumber
    .
\eeq
Re-expressing the $G$ superoperators in terms of $J$ superoperators and tunnel amplitudes, we have
\beq
  \Sigma^{(2)}(z) &=&  
  - p_1 p_2  
  J^{p_2}_{\bar{1} m_2} \opp{\phi_a}{\phi_a} J^{p_1}_{1 m_1}
  \nonumber\\
  &&
  \times
  t_{\bar{1}m_2} t_{1 m_1}
  \frac{ f_{\alpha_1}(-\xi_1 p_1 \omega_1)}
    {z+i\xi_1(\omega_1+\mu_{\alpha_1}) +i\Delta_a}
  .
  \nonumber  \\
  \label{SIG2tt}
\eeq
With rates defined as in \eq{rates} with the constant tunnelling density of states approximation, \eq{SIG2tt} becomes
\beq
 \Sigma^{(2)}(z) &=&  
  - p_1 p_2  
  J^{p_2}_{\bar{1} m_2} \opp{\phi_a}{\phi_a} J^{p_1}_{1 m_1}
  \nonumber\\
  &&
  \times
   \Gamma_{\xi_1 \alpha_1}^{ m_2 m_1}
   I_a^{(2)}(z;\xi_1,p_1,\mu_1)
   ,
\eeq
with integral
\beq
   I_a^{(2)} =  
   \frac{1}{2\pi}
   \int_{-D}^{D} d\omega_1
   \frac{f_{\alpha_1}(-\xi_1 p_1 \omega_1)}
     { z+i\xi_1(\omega_1+\mu_{\alpha_1}) +i\Delta_a} 
   \nonumber
   ,
\eeq
where the bandwidth $D$ is assumed to be much larger than all other energy scales in the problem.
We regularise the integrals by setting $z\to 0^+ -i \epsilon$ with $\epsilon$ wholly real.  We are therefore left to evaluate the integral
\beq
   I_a^{(2)} =
   \frac{1}{2\pi}
   \int_{-D}^{D} d\omega_1
   \frac{f_{\alpha_1}(-\xi_1 p_1 \omega_1)}
     {i0^+ -i \epsilon +i\xi_1(\omega_1+\mu_{\alpha_1}) +i\Delta_a} 
   \nonumber
   .
\eeq
Use of Dirac's identity, $(i0^+ + x)^{-1} = {\cal P}[1/x]-i\pi \delta(x)$, allows us to write
\beq
  I_a^{(2)}(z=0^+ - i \epsilon) &=&  
  \frac{1}{2} f_{\alpha_1}( p_1(\Delta_a + \xi_1\mu_{\alpha_1}-\epsilon))
  \nonumber\\
  &&
  +\frac{i p_1}{2\pi}\phi_{\alpha_1}(p_1(\Delta_a + \xi_1\mu_{\alpha_1}-\epsilon))
  \nonumber
  ,
\eeq
with $\phi$ defined in \eq{phidef}, correct to order $1/D$.

\subsection{Fourth order}

At fourth order, there are two linked contractions, $(41)(32)$ and $(42)(31)$, we we label ``D'' for direct and ``X'' for exchange.  The direct contribution reads
\beq
  \Sigma^{(4D)}(z) &=& 
  G^{p_4}_{\bar{1}}  \Omega_\mathrm{S}(z_3)
  G^{p_3}_{\bar{2}}  \Omega_\mathrm{S}(z_2)
  \nonumber\\
  &&\times
  G^{p_2}_2 \Omega_\mathrm{S}(z_1)
  G^{p_1}_1
  \gamma_{41}^{p_4 p_1} \gamma_{32}^{p_3 p_2}
  .
\eeq
This evaluates as \eq{SDmain} with the integral 
\beq
  I^{D} &=& \frac{1}{(2\pi)^2}
  \frac{-i}{\lambda_3 - \lambda_1} \int d\omega_1\int d\omega_2
  \nonumber\\
  && \times
  \frac
  {f(p_1 \omega_1)f( p_2 \omega_2)}
  {
    (i0^+ + \omega_1 + \omega_2 - \lambda_2)
    (i0^+ + \omega_1 - \lambda_3)
  } 
  \nonumber\\
  && +
  \rb{\lambda_1 \leftrightarrow \lambda_3}
\eeq
with
$\lambda_1 = \xi_1 \mu_{\alpha_1} + \Delta_{a''}-\epsilon$,
$\lambda_2 = \xi_1 \mu_{\alpha_1} + \xi_2 \mu_{\alpha_2} + \Delta_{a'}-\epsilon$,
$\lambda_3 = \xi_1 \mu_{\alpha_1} + \Delta_{a}-\epsilon$.  Note that here the $\rb{\lambda_1 \leftrightarrow \lambda_3}$ symbol includes the pre-integral forefactor.

The exchange term is
\beq
  \Sigma^{(4X)}(z) &=& 
  -
  G^{p_4}_{\bar{2}}  \Omega_\mathrm{S}(z_3)
  G^{p_3}_{\bar{1}}  \Omega_\mathrm{S}(z_2)
  \nonumber\\
  &&
  \times
  G^{p_2}_2  \Omega_\mathrm{S}(z_1)
  G^{p_1}_1
   \gamma_{42}^{p_4 p_2} \gamma_{31}^{p_3 p_1}
   ,
\eeq
which yields \eq{SXmain} with exchange integral
\beq
  I^{X}
  &=&
  \frac{1}{(2\pi)^2}
  \frac{-i}{\lambda_2-\lambda_3 - \lambda_1} 
  \int d\omega_1\int d\omega_2
  f(p_1 \omega_1)f( p_2 \omega_2) 
  \nonumber\\
  &&\times
  \rb{
    \frac{1}{i0^+  + \omega_1 - \lambda_1}
    +
    \frac{1}{i0^+  + \omega_2 - \lambda_3}
  }
  \nonumber\\
  && 
  \times
  \rb{
  \frac{1}{i0^+ + \omega_1 + \omega_2 - \lambda_2}
  \right.
  \nonumber\\
  &&
  \left.
  ~~~~~~~~~~~~~~~~~
  -
  \frac{1}{i0^+  + \omega_1 + \omega_2 - \lambda_1-\lambda_3}
  }
  ,
  \nonumber\\
\eeq
with
$\lambda_1 = \xi_1 \mu_{\alpha_1} + \Delta_{a''}-\epsilon$,
$\lambda_2 = \xi_1 \mu_{\alpha_1} + \xi_2 \mu_{\alpha_2} + \Delta_{a'}- \epsilon$, and
$\lambda_3 = \xi_2 \mu_{\alpha_2} + \Delta_{a}-\epsilon$.

The analytic evaluation of these integrals is discussed in Ref.~\cite{lei08}.  Let us note here that Dirac's identity can be used to split these integrals up in three parts, e.g. $I^D = I^{D0}+ I^{D1}+ I^{D2}$, where the number in the superscript  corresponds to the number of delta functions appearing in the integrand.  Integrals with a single delta function give the cotunneling rates and are the most important for the current scheme.  These are evaluated as
\beq
   I^{D1}
   &=& 
  \frac{p_1 p_2 }{(2\pi)^2}
  \frac{
    F (\lambda_2,\lambda_3)  -  F (\lambda_2,\lambda_1) 
  }{\lambda_3-\lambda_1}
  \nonumber\\
  &&+
  \frac{p_1}{(2\pi)^2}
  \frac{
    \widetilde{F}(\lambda_3)  - \widetilde{F} (\lambda_1) 
  }{\lambda_3-\lambda_1}
\eeq
with
\beq
 F (\lambda',\lambda) &=&
 -\pi
 \rb{ \frac{}{}
   - b(\lambda') 
   \left[
     \phi(-\lambda) - \phi(\lambda'-\lambda)
   \right]
  \right.
  \nonumber\\
  &&
  \left.
   -\frac{1}{2} \phi(\lambda)
   +\phi(\lambda'-\lambda)f(\lambda)
 }
 ,
\eeq 
$\widetilde{F}(\lambda) = -\frac{\pi}{2} \phi(\lambda)$, and where $b(x) = \rb{e^{x/T}-1}^{-1}$ is the Bose-Einstein distribution.
The single delta-function Exchange integral gives
\beq
  I^{X1} 
  &=&
  \frac{ p_1 p_2}{(2\pi)^2}
  \rb{
  \frac{
    F(\lambda_2,\lambda_1)
    - F(\lambda_1+\lambda_3,\lambda_1)
  }{\lambda_2-\lambda_3-\lambda_1}
  \right.
  \nonumber\\
  &&+
  \left.
  \frac{
    F(\lambda_2,\lambda_3)
    - F(\lambda_1+\lambda_3,\lambda_3)
   }{\lambda_2-\lambda_3-\lambda_1}
  }
  .
\eeq
The remaining parts give rise to renormalisation of the system levels, and whereas  $I^{D2}$ and $I^{X2}$ can easily be evaluated analytically, the double principal-part integrals must be performed numerically.  In the models studies here, however, these fourth-order renormalisation parts were unimportant.

\section{First three cumulants in scattering approach \label{appSCAT}}

In the two-terminal scattering formalism \cite{but92,bla00}, the average current and shotnoise at finite temperature and bias are given by the well-known expressions 
\beq
  I &=& 
  \frac{1}{2\pi}
  \int d E ~
  T(E)
  \left[
   f_L(E)
    -
   f_R(E)
  \right]
  \nonumber \\
  S &=&  \frac{1}{2\pi}
  \int dE
  \left\{\frac{}{}
    T(E) [f_L(1-f_L) + f_R(1-f_R)]
  \right.
  \nonumber\\
  &&~~~~~~~~~~~~~~~~~~~
  \left.
    +
     T(E) \rb{1- T(E)} (f_L-f_R)^2
  \frac{}{}\right\}
  \nonumber
  ,
\eeq
where $T(E)$ is the transmission probability of the device and $f_X$ is the Fermi function of lead $X$.  We have set here $e=\hbar=1$ and define the correlation functions in agreement with those of FCS.  The corresponding expression for the skewness at finite temperature and bias is less well known.  However, from the results for the symmetrised correlator of \cite{sal06}  (the appropriate quantity here), we have
\beq
   S^{(3)}=S^{(3)}_\mathrm{SYM.} = 
  \frac{1}{2\pi} \int dE 
  \left\{ 
    3 S_{ioo}(E) - S_{ooo}(E)
  \right\}
\eeq
with
\beq
  S_{ioo} &=& (1 - T)^2 f_L(1 - f_L)(1 - 2f_L)  
  \nonumber\\
  &&
  + T(1 - T)f_L(1 - f_L)(1 - 2f_R) 
  \nonumber\\
  S_{ooo} &=& 
  (1-T)^3 f_L (1 - f_L) (1 - 2f_L) 
  + T(1 - T)^2 a_{LR}
  \nonumber\\
  &&
  + T^2 (1 - T) a_{RL} 
  + T^3 f_R(1 - f_R)(1 - 2f_R)
  \nonumber
  ,
\eeq
and
\beq
  a_{XY} &=& f_X(1 - f_X)(1 - 2 f_Y) + f_X(1 - f_Y)(1 - 2f_X) 
  \nonumber\\
  &&
  + f_Y(1 - f_X)(1 - 2f_X).
\eeq
In the infinite bias limit, $f_L=1$ and $f_R=0$, we recover
\beq
  S^{(3)}_\mathrm{SYM.} = 
  \frac{1}{2\pi} \int dE 
  \left\{ 
    T(E)(1-T(E))(1-2T(E))
  \right\} 
  \nonumber
  ,
\eeq
which is the more familiar expression for skewness in the scattering approach.

From Ref.~\cite{bla00} we know that the transmission probability of the single resonant level at energy $\epsilon$ with partial widths $\Gamma_L$ and $\Gamma_R$ is
\beq
  T(E) =
  \frac{\Gamma_L \Gamma_R}
  {(E -\epsilon)^2 + (\Gamma_L+ \Gamma_R)^2/4}.
\eeq
\begin{acknowledgments}
Work supported by the WE Heraeus foundation and by DFG grant BR 1528/5-1.  I am grateful to M.~Wegewijs and M.~Leijnse for extensive discussions and to T.~Brandes, R.~Aguado, M.~Hettler, and B.~W\"unsch for their helpful input.
\end{acknowledgments}



\begin{thebibliography}{99}
\bibitem{FN_nonMark}  In general, a nonMarkovian master equation also includes an inhomogeneous term $\gamma(t)$. However, as we restrict ourselves to zero-frequency current correlations,  $\gamma(t)$ can be neglected \cite{fli08}.
\bibitem{FNinf}
  In the infinite bias limit the sequential-tunnelling effective Liouvillian becomes exact \cite{gur96}, and thus cumulants of arbitrary order may safely be considered.
\end{thebibliography}
\end{document}